\documentclass[a4paper,10pt]{article}
\usepackage[dvips]{graphicx}
\usepackage{amssymb,amsmath}
\numberwithin{equation}{section}

\begin{document}

\title{Some cosmological aspects of  Ho$\check{r}$ava-Lifshitz  gravity: integrable and nonintegrable  models}
\vspace{4cm}
\author{ G.N. Nugmanova$^1$,  Sh.R. Myrzakul$^1$, O.V.  Razina$^1$, K.R.  Esmakhanova$^1$,   \\N.S.  Serikbayev$^1$,    R. Myrzakulov$^{1,2}$\footnote{The corresponding author. Email: rmyrzakulov@gmail.com; rmyrzakulov@csufresno.edu}\\ \textit{$^1$Eurasian International Center for Theoretical Physics,} \\ \textit{Eurasian National University, Astana 010008, Kazakhstan}\\ \textit{$^2$Department of Physics, CSU Fresno, Fresno, CA 93740 USA}}

\date{}

\maketitle
\begin{abstract} 

In this work, some new integrable and nonintegrable cosmological models of the  Ho$\check{r}$ava-Lifshitz  gravity  are proposed. For some of them, exact solutions are presented. Then these results extend for the F(R) Ho$\check{r}$ava-Lifshitz  gravity theory case. In particular, several integrable cosmological models of this modified gravity theory were constructed in the explicit form. 
\end{abstract}
\sloppy
\section{Introduction} 
 More than one year ago Ho$\check{r}$ava proposed a theory, the so-called  Ho$\check{r}$ava-Lifshitz quantum gravity, which is a power-counting renormalizable theory with consistent ultra-violet behavior \cite{Ho}. In this theory, the scaling of the gravitational system at short distances exhibits a strong anisotropy between space and time 
 \begin{equation}
x^i\rightarrow bx^i, \quad t\rightarrow b^zt. \end{equation}
This important relation between space and time coordinates can be realized with the Arnowitt-Deser-Misner decomposition of the metric of the form
\begin{equation}
ds^2=-L^2dt^2+g_{ij}(dx^i+L^idt)(dx^j+L^jdt), \end{equation}
where $g_{ij}$ is the spatial metric (roman letters indicate spatial indices), $L$ and $L_{i}$ are the lapse and shift functions, respectively. Recently the F(R) Ho$\check{r}$ava-Lifshitz quantum gravity has been proposed \cite{Chaichian} (see also \cite{Odin1}-\cite{Saez2}).  In this work, we consider integrable aspects of the usual and F(R) modified  Ho$\check{r}$ava-Lifshitz  gravity theories ( see e.g. \cite{Ablowitz}-\cite{MR}).

The paper is organized as follows. In section 2, we  study
some integrable and nonintegrable FRW cosmological models of Ho$\check{r}$ava-Lifshitz  gravity. In the next section 3, we extend some of these results  for the modified F(R) Ho$\check{r}$ava-Lifshitz quantum gravity case and present its some exact integrable models. Exact solutions of some integrable models of the usual  Ho$\check{r}$ava-Lifshitz  gravity were considered in the section 4. Section 5 is  devoted to the conclusion.
\section{Cosmological models of Ho$\check{r}$ava-Lifshitz gravity}

In this work, we restrict ourselves to  the detailed balance case. In this case, the action of Ho$\check{r}$ava-Lifshitz gravity takes the form ($z=3$):
$$
S=\int dtd^3x\sqrt{g}N[\frac{2}{\kappa^2}(K_{ij}K^{ij}-\lambda K^2)+\frac{\kappa^2}{2w^4}C_{ij}C^{ij}-\frac{\kappa^2\mu\epsilon^{ijk}}{2w^2\sqrt{g}}R_{il}\nabla_{j}R^{l}_{k}+$$
\begin{equation}
\frac{\kappa^2\mu^2}{8}R_{ij}R^{ij}-\frac{\kappa^2\mu^2}{8(3\lambda-1)} (\frac{1-4\lambda}{4}R^2+\Lambda R-3\Lambda^2)], \end{equation}
where $\nabla_i$ are  the covariant derivatives defined with respect to the spatial metric $g_{ij}$, $\epsilon^{ijk}$ is the totally antisymmetric unit tensor and  $\kappa, \lambda, \mu, w = consts $. Here
$K_{ij}$ and $C^{ij}$ are the extrinsic curvature and the Cotton tensor, respectively, which are given by
\begin{equation}
K_{ij}=0.5L^{-1}(\dot{g}_{ij}-\nabla_iL_j-\nabla_jL_i),\quad C^{ij}=g^{-0.5}\epsilon^{ijk}\nabla_k(R^j_i-0.25R\delta^j_i). \end{equation}
As our main interest in this work  is the cosmological aspects of HL gravity, we impose the projectability  condition.  Consider FRW spacetime  with the scale factor $a(t)$ and metric
   \begin{equation}
ds^2=-dt^2+a^2(t)\left[\frac{dr^2}{1-kr^2}+r^2d\Omega^2\right] \end{equation}
that is
 \begin{equation}
L=1, \quad g_{ij}=a^2(t)\gamma_{ij}, \quad L^i=0, \end{equation}
where  $k$ can take any value but it is related to $(-, 0, +)$ curvatures according to sign.  In this case, the Friedmann equations  we can write\\ in   \textit{the H-form}
	\begin{eqnarray}
		p&=&-\alpha(2\dot{H}+3H^2)+f_1(a),\\
	\rho&=&3\alpha H^2+f_2(a),\\
	\dot{\rho}&=&-3H(\rho+p)
	\end{eqnarray}
	or in  \textit{the N-form}	\begin{eqnarray}
	p&=&-\alpha(2\ddot{N}+3\dot{N}^2)+f_1(a),\\
	\rho&=&3\alpha\dot{N}^2+f_2(a),\\
	\dot{\rho}&=&-3\dot{N}(\rho+p).
	\end{eqnarray}
Here $N=\ln{a}, \quad H=\dot{N}
$ and
 \begin{equation}
f_1=\beta_0+\beta_{2}a^{-2}+\beta_{4}a^{-4}, \quad f_2=\eta_0+\eta_2a^{-2}+\eta_4a^{-4}, \end{equation}
where $\alpha=2\kappa^{-2}(3\lambda-1)$ and
\begin{equation}
\beta_0=0.75\mu^2\Lambda^2\alpha^{-1},\quad \beta_2=-0.5k\mu^2\Lambda\alpha^{-1},\quad  \beta_4=-0.25 \mu^2k^2\alpha^{-1},\end{equation}
\begin{equation}
\eta_0=-0.75\mu^2\Lambda^2\alpha^{-1}=-\beta_0,\quad \eta_2=1.5k\mu^2\Lambda\alpha^{-1}=-3\beta_2,\quad  \eta_4=0.75 \mu^2k^2\alpha^{-1}=-3\beta_4.\end{equation}
In the next sections we will study some integrable aspects of the Friedmann equations in the H-form (2.5)-(2.7) or in the N-form (2.8)-(2.10) and try solve some of them.

\subsection{Integrable  models}
Let us start from the  presentation of   some examples of integrable HL cosmological models. 
 We  assume that $N, a, H$ satisfy one of Painlev$\acute{e}$
	equations so that we obtain 30 new integrable HL models. Consider examples [below $\alpha, \beta, \gamma, \delta, \kappa$ and $\mu$ are arbitrary constants] (see also \cite{Ablowitz}-\cite{MR}). 
	
1) {\bf P$_{I}$ - models.} In our cosmological case, this model as and  other P-type models have  5 particular submodels that means 5 type cosmological models. For this case we have:
\\
i) P$_{IA}$ - model. The P$_{IA}$ - model we  write in the following closed form
\begin{eqnarray}
		p&=&-\alpha(2\dot{H}+3H^2)+f_1(a),\\
	\rho&=&3\alpha H^2+f_2(a),\\
		\ddot{N}&=&6N^2+t,\\\dot{\rho}&=&-3H(\rho+p)
		\end{eqnarray}
	or
		\begin{eqnarray}
	p&=&-\alpha(2\ddot{N}+3\dot{N}^2)+f_1(a),\\
	\rho&=&3\alpha\dot{N}^2+f_2(a),\\
	\ddot{N}&=&6N^2+t, \\
		\dot{\rho}&=&-3\dot{N}(\rho+p).
	\end{eqnarray}
We believe that the system (2.14)-(2.17) [or its equivalent (2.18)-(2.21)] is integrable. 	Note that the  case  $\alpha=1, \quad f_i=0$ corresponds to General Relativity (GR).  Keeping in mind that the P$_{IA}$ - model has the form (2.14)-(2.17) [or equivalently (2.18)-(2.21)], for short, we here write it as
	\begin{equation}
 \ddot{N}=6N^2+t.
 \end{equation}
 We also note that in the systems (2.14)-(2.17) or (2.18)-(2.21), the equation (2.22)  plays the role of the  EoS $p=p(\rho)$. Finally we would like to note that similarly the other models we below write in the same short form as (2.22). \\
 ii) P$_{IB}$ - model:
\begin{equation}
 \ddot{a}=6a^2+t.\end{equation}
  iii) P$_{IC}$ - model:
\begin{equation}
 \ddot{H}=6H^2+t.\end{equation}
  iv)  P$_{ID}$ - model:
\begin{equation}
 H_{aa}=6H^2+a.\end{equation}
  v)  P$_{IE}$ - model:
\begin{equation}
 H_{NN}=6H^2+N.\end{equation}
  
2) {\bf P$_{II}$ - models.}
\\
i) P$_{IIA}$ - model:\begin{equation}
 \ddot{N}=2N^3+tN+\nu.\end{equation}
ii)  P$_{IIB}$ - model:
 \begin{equation}
 \ddot{a}=2a^3+ta+\nu.\end{equation}
 iii)  P$_{IIC}$ - model:
 \begin{equation}
 \ddot{H}=2H^3+tH+\nu.
 \end{equation}
  iv)  P$_{IID}$ - model:
 \begin{equation}
 H_{aa}=2H^3+aH+\nu.
 \end{equation}
   v)  P$_{IIE}$ - model:
 \begin{equation}
 H_{NN}=2H^3+NH+\nu.
 \end{equation}
 3) {\bf P$_{III}$ - models.}
\\
i) P$_{IIIA}$ - model:
\begin{equation}
 \ddot{N}=\frac{1}{N}\dot{N}^2-\frac{1}{t}(\dot{N}-\alpha N^2-\beta)+\gamma N^3+\frac{\delta}{N}.\end{equation}
ii) P$_{IIIB}$ - model:
\begin{equation}
 \ddot{a}=\frac{1}{a}\dot{a}^2-\frac{1}{t}(\dot{a}-\alpha a^2-\beta)+\gamma a^3+\frac{\delta}{a}.\end{equation}
 iii) P$_{IIIC}$ - model:
\begin{equation}
 \ddot{H}=\frac{1}{H}\dot{H}^2-\frac{1}{t}(\dot{H}-\alpha H^2-\beta)+\gamma H^3+\frac{\delta}{H}.\end{equation}
  iv) P$_{IIID}$ - model:
\begin{equation}
 H_{aa}=\frac{1}{H}H_{a}^2-\frac{1}{a}(H_{a}-\alpha H^2-\beta)+\gamma H^3+\frac{\delta}{H}.\end{equation}
   v) P$_{IIIE}$ - model:
\begin{equation}
 H_{NN}=\frac{1}{H}H_{N}^2-\frac{1}{N}(H_{n}-\alpha H^2-\beta)+\gamma H^3+\frac{\delta}{H}.\end{equation}
 
 4) {\bf P$_{IV}$ - models. }
 \\
i)  P$_{IVA}$ - model:
\begin{equation}
 \ddot{N}=\frac{1}{2N}\dot{N}^2+1.5N^3+4tN^2+2(t^2-\alpha)N+
 \frac{\delta}{N}.\end{equation}
 ii)  P$_{IVB}$ - model:
\begin{equation}
 \ddot{a}=\frac{1}{2a}\dot{a}^2+1.5a^3+4ta^2+2(t^2-\alpha)a+
 \frac{\delta}{a}.\end{equation}
 iii)  P$_{IVC}$ - model:
\begin{equation}
 \ddot{H}=\frac{1}{2H}\dot{H}^2+1.5H^3+4tH^2+2(t^2-\alpha)H+
 \frac{\delta}{H}.\end{equation}
 iv)  P$_{IVD}$ - model:
\begin{equation}
 H_{aa}=\frac{1}{2H}H_{a}^2+1.5H^3+4aH^2+2(a^2-\alpha)H+
 \frac{\delta}{H}.\end{equation}
   v)  P$_{IVE}$ - model:
\begin{equation}
 H_{NN}=\frac{1}{2H}H_{N}^2+1.5H^3+4NH^2+2(N^2-\alpha)H+
 \frac{\delta}{H}.\end{equation}
  5) {\bf P$_{V}$ - models. }
  \\
 i) P$_{VA}$ - model: 
  \begin{equation}
 \ddot{N}=(\frac{1}{2N}+\frac{1}{N-1})\dot{N}^2
 -\frac{1}{t}(\dot{N}-\gamma  N)+t^{-2}(N-1)^2(\alpha N+\beta N^{-1})
 +\frac{\delta N(N+1)}{N-1}.\end{equation}
 ii) P$_{VB}$ - model: 
  \begin{equation}
 \ddot{a}=(\frac{1}{2a}+\frac{1}{a-1})\dot{a}^2
 -\frac{1}{t}(\dot{a}-\gamma  a)+t^{-2}(a-1)^2(\alpha a+\beta a^{-1})
 +\frac{\delta a(a+1)}{a-1}.\end{equation}
  iii) P$_{VC}$ - model: 
  \begin{equation}
 \ddot{H}=(\frac{1}{2H}+\frac{1}{H-1})\dot{H}^2
 -\frac{1}{t}(\dot{H}-\gamma  H)+t^{-2}(H-1)^2(\alpha H+\beta H^{-1})
 +\frac{\delta H(H+1)}{H-1}.\end{equation}
  iv) P$_{VD}$ - model: 
  \begin{equation}
 H_{aa}=(\frac{1}{2H}+\frac{1}{H-1})H_{a}^2
 -\frac{1}{a}(H_{a}-\gamma  H)+a^{-2}(H-1)^2(\alpha H+\beta H^{-1})
 +\frac{\delta H(H+1)}{H-1}.\end{equation}
  v) P$_{VE}$ - model: 
  \begin{equation}
 H_{NN}=(\frac{1}{2H}+\frac{1}{H-1})H_{N}^2
 -\frac{1}{N}(H_{N}-\gamma  H)+N^{-2}(H-1)^2(\alpha H+\beta H^{-1})
 +\frac{\delta H(H+1)}{H-1}.\end{equation}
 6) {\bf P$_{VI}$ - models. }
 \\
 i) P$_{VIA}$ - model:
$$
 \ddot{N}=0.5\left(\frac{1}{N}+\frac{1}{N-1}+\frac{1}{N-t}\right)\dot{N}^2
 -\left(\frac{1}{t}+\frac{1}{t-1}+\frac{1}{N-t}\right)\dot{N}
 $$\begin{equation}+t^{-2}(t-1)^{-2}N(N-1)(N-t)\left[\alpha+ \beta tN^{-2}+\gamma(t-1)(N-1)^{-2}+\delta t(t-1)(N-t)^{-2}\right].\end{equation}
 ii) P$_{VIB}$ - model:
$$
 \ddot{a}=0.5\left(\frac{1}{a}+\frac{1}{a-1}+\frac{1}{a-t}\right)\dot{a}^2
 -\left(\frac{1}{t}+\frac{1}{t-1}+\frac{1}{a-t}\right)\dot{a}
 $$\begin{equation}+t^{-2}(t-1)^{-2}a(a-1)(a-t)\left[\alpha+ \beta ta^{-2}+\gamma(t-1)(a-1)^{-2}+\delta t(t-1)(a-t)^{-2}\right].\end{equation}
 iii) P$_{VIC}$ - model:
$$
 \ddot{H}=0.5\left(\frac{1}{H}+\frac{1}{H-1}+\frac{1}{H-t}\right)\dot{H}^2
 -\left(\frac{1}{t}+\frac{1}{t-1}+\frac{1}{H-t}\right)\dot{H}
 $$\begin{equation}+t^{-2}(t-1)^{-2}H(H-1)(H-t)\left[\alpha+ \beta tH^{-2}+\gamma(t-1)(H-1)^{-2}+\delta t(t-1)(H-t)^{-2}\right].\end{equation}
iv) P$_{VID}$ - model:
$$
 H_{aa}=0.5\left(\frac{1}{H}+\frac{1}{H-1}+\frac{1}{H-a}\right)H_{a}^2
 -\left(\frac{1}{a}+\frac{1}{a-1}+\frac{1}{H-a}\right)H_{a}
 $$\begin{equation}+a^{-2}(a-1)^{-2}H(H-1)(H-a)\left[\alpha+ \beta aH^{-2}+\gamma(a-1)(H-1)^{-2}+\delta a(a-1)(H-a)^{-2}\right].\end{equation}
v) P$_{VIE}$ - model:
$$
 H_{NN}=0.5\left(\frac{1}{H}+\frac{1}{H-1}+\frac{1}{H-N}\right)H_{N}^2
 -\left(\frac{1}{N}+\frac{1}{N-1}+\frac{1}{H-N}\right)H_{N}
 $$\begin{equation}+N^{-2}(N-1)^{-2}H(H-1)(H-N)\left[\alpha+ \beta NH^{-2}+\gamma(N-1)(H-1)^{-2}+\delta N(N-1)(H-N)^{-2}\right].\end{equation}

Note  that all Painlev$\acute{e}$ equations  can be represented as Hamiltonian systems that is as (see e.g. \cite{Ablowitz} and references therein)
 \begin{eqnarray}
		\dot{q}&=&\frac{\partial E}{\partial r},\\
	\dot{r}&=&-\frac{\partial E}{\partial q},
	\end{eqnarray}
 where $E(q, r, t)$ is the (non-autonomous) Hamiltonian function. Consider some examples (see e.g. \cite{Ablowitz} and references therein).
 
 1) P$_{I}$-models.  In this case, $q,r,F$ read as 
 \begin{eqnarray}
		\dot{q}&=&r,\\
	\dot{r}&=&6q^2+t,\\
	E&=&0.5r^2-2q^3-tq.
	\end{eqnarray}
	
	 2) P$_{II}$-models.  In this case we have
 \begin{eqnarray}
		\dot{q}&=&r-q^2-0.5t,\\
	\dot{r}&=&2qr+\alpha+0.5,\\
	E&=&0.5r^2-(q^2+0.5t)r-(\alpha+0.5)q.
	\end{eqnarray}
	
	 3) P$_{III}$-models.  In this case we have
 \begin{eqnarray}
		t\dot{q}&=&2q^2r-k_2tq^2-(2\theta_1+1)q+k_1t,\\
	t\dot{r}&=&-2qr^2+2k_2tqr+(2\theta_1+1)r-k_2((\theta_1+\theta_2)t,\\
tE&=&q^2r^2-[k_2tq^2+(2\theta_1+1)q-k_1t]r+k_2(\theta_1+\theta_2)tq.
	\end{eqnarray}
	So in this subsection we presented 24 new HL cosmological models. Note that as integrable systems,  these models admit n-soliton solutions, infinite number commuting integrals of motion, Lax representations etc. 
\subsection{Nonintegrable models}
Here we consider   some known and new  HL models induced by  some  ODEs. These ODEs are   nonintegrable so that the corresponding HL 
cosmological models are nonintegrable (see also \cite{MR}).
 
1) {\bf $\Lambda$CDM cosmology}. 
We start with the $\Lambda$CDM cosmology. Here we present 5 submodels [We remark that in fact only the  $\Lambda_1$ - model corresponds  to $\Lambda$CDM cosmology].
\\
i)  $\Lambda_1$ - model:
\begin{equation}
 \ddot{N}=0.5\Lambda-1.5\dot{N}^2.\end{equation}
 ii)  $\Lambda_2$ - model:
\begin{equation}
 \ddot{a}=0.5\Lambda-1.5\dot{a}^2.\end{equation}
 iii)  $\Lambda_3$ - model:
\begin{equation}
 \ddot{H}=0.5\Lambda-1.5\dot{H}^2.\end{equation}
   iv)  $\Lambda_4$ - model:
\begin{equation}
 H_{aa}=0.5\Lambda-1.5H_{a}^2.\end{equation}
   v)  $\Lambda_5$ - model:
\begin{equation}
 H_{NN}=0.5\Lambda-1.5H_{N}^2.\end{equation}
 
 2) {\bf Pinney  cosmology}.  It induced by  the Pinney equation. Let us present 5 submodels.
 \\
 i) Pinney$_1$ - model:
 \begin{equation}
 \ddot{N}=\xi(t)N+\frac{k}{N^3},\end{equation}
 where $\xi=\xi(t).\quad k=const$.
 \\
ii) Pinney$_2$ - model:
 \begin{equation}
 \ddot{a}=\xi(t)a+\frac{k}{a^3}.\end{equation}
  \\
iii) Pinney$_3$ - model:
 \begin{equation}
 \ddot{H}=\xi(t)H+\frac{k}{H^3}.\end{equation}
   \\
iv) Pinney$_4$ - model:
 \begin{equation}
 H_{aa}=\xi(a)H+\frac{k}{H^3}.\end{equation}
  \\
v) Pinney$_5$ - model:
 \begin{equation}
 H_{NN}=\xi(N)H+\frac{k}{H^3}.\end{equation}
 
3) {\bf Schr$\ddot{o}$dinger cosmology}.  For this model also write 5 submodels. 
\\
i) Schr$\ddot{o}$dinger$_1$ - model:
 \begin{equation}
 \ddot{N}=uN+kN,\end{equation}
 where $u=u(t),\quad k=const$. 
 \\
ii) Schr$\ddot{o}$dinger$_2$ - model:
 \begin{equation}
 \ddot{a}=ua+ka.\end{equation}
\\
iii) Schr$\ddot{o}$dinger$_3$ - model:
 \begin{equation}
 \ddot{H}=uH+kH.\end{equation}
 \\
iv) Schr$\ddot{o}$dinger$_4$ - model:
 \begin{equation}
 H_{aa}=uH+kH.\end{equation}
  \\
v) Schr$\ddot{o}$dinger$_5$ - model:
 \begin{equation}
 H_{NN}=uH+kH.\end{equation}
 
 4) {\bf Hypergeometric cosmology}. Let us we present  5 submodels.
  \\ 
  i) H$_1$ - model:
  \begin{equation}
 \ddot{N}=t^{-1}(1-t)^{-1}\{[(\alpha+\beta+1)t-\gamma]\dot{N}+\alpha\beta N\}.\end{equation}
 \\
 ii) H$_2$ - model:
  \begin{equation}
 \ddot{a}=t^{-1}(1-t)^{-1}\{[(\alpha+\beta+1)t-\gamma]\dot{a}+\alpha\beta a\}.\end{equation}
 \\
  iii) H$_3$ - model:
  \begin{equation}
 \ddot{H}=t^{-1}(1-t)^{-1}\{[(\alpha+\beta+1)t-\gamma]\dot{H}+\alpha\beta H\}.\end{equation}
\\
  iv) H$_4$ - model:
  \begin{equation}
 H_{aa}=a^{-1}(1-a)^{-1}\{[(\alpha+\beta+1)a-\gamma]H_{a}+\alpha\beta H\}.\end{equation}
\\
  v) H$_5$ - model:
  \begin{equation}
 H_{NN}=N^{-1}(1-N)^{-1}\{[(\alpha+\beta+1)N-\gamma]H_{N}+\alpha\beta H\}.\end{equation}

	Finally we would like to note that the above presented HL models, in particular, have some solutions which describe the accelerated expansion of the universe. 
 \section{Cosmological models of F(R) Ho$\check{r}$ava-Lifshitz gravity}
 Let us now consider the F(R) Ho$\check{r}$ava-Lifshitz gravity. Following \cite{Chaichian}, its action we write as
 \begin{equation}
S=0.5\kappa^{-2}\int dtd^3x\sqrt{g}NF(\tilde{R}),
 \end{equation}
where $\kappa^2=16\pi G$ and 
 \begin{equation}
\tilde{R}=K_{ij}K^{ij}-\lambda K^2+R+2\mu \nabla_{\mu}(n^{\mu}\nabla_{\nu}n^{\nu}-n^{\nu}\nabla_{\nu}n^{\mu})-L^{(3)}g_{ij}.
 \end{equation}
 The case $F(\tilde{R})=\tilde{R}$ corresponds to the original Ho$\check{r}$ava-Lifshitz gravity. In this section, we construct   integrable and  nonintegrable F(R) Ho$\check{r}$ava-Lifshitz gravity models induced by  some ODSs. For the metric 
  \begin{equation}
ds^2=-L^2dt^2+a^2(t)(dx^2+dy^2+dz^2),
 \end{equation} the scalar $\tilde{R}$ takes the form
  \begin{equation}
\tilde{R}=3(1-3\lambda+6\mu)H^2L^{-2}+6\mu H^{-1}[\ln(HL^{-1})]_{t}.
 \end{equation}
 In the FRW case and as $L=1$, the equations of motion read as \cite{Chaichian}
  \begin{eqnarray}
0&=&F-2(1-3\lambda+3\mu)(\dot{H}+3H^2)F^{'}-2(1-3\lambda)\dot{\tilde{R}}F^{''}+2\mu\dot{\tilde{R}}^2F^{'''}+\ddot{\tilde{R}}F^{''})+\kappa^2 p_m,\\
0&=&F-6[(1-3\lambda+3\mu)H^2+\mu\dot{H}]F^{'}+6\mu H\dot{\tilde{R}}F^{''}-\kappa^2 \rho_m,
	\end{eqnarray}
where   $p_m$ and $\rho_m$ are the pressure and energy density of a perfect fluid that fills the Universe.
 \subsection{Integrable  models}
 Some classes  integrable models  can be constructed with the help of known integrable ODEs. It is the main idea of the work (see also \cite{MR}). One of famous representatives of such integrable ODEs are  Painlev$\acute{e}$ equations. It is the case that we are going to use to get integrable Ho$\check{r}$ava-Lifshitz gravity models. Let us demonstrate it. The lazy way to do it is the   assumption   that the function $F(\tilde{R})$ or its sister  $f(\tilde{R})=F(\tilde{R})-\tilde{R}$  satisfy some integrable ODEs,  in our case,  one of Painlev$\acute{e}$ equations.  Note that there are 6 integrable Painlev$\acute{e}$ equations with each of them can be  relate 4 integrable $F(\tilde{R})$  Ho$\check{r}$ava-Lifshitz models so that totally we have 30 models. Now we are in the position to  present these  equations (or models) [$F^{'}\equiv dF/d\tilde{R}, \quad \dot{F}\equiv dF/dt$ etc]. 
 \\
 1) {\bf  $F_{I}(R)$ - models.}
 \\a) $F_{IA}(R)$ - model:
  \begin{equation} 
 f^{''}=6f^2+\tilde{R}\end{equation}
 or
   \begin{equation}
 F^{''}=6F^2+\tilde{R}.\end{equation}
 b) $F_{IB}(R)$ - model:
   \begin{equation}
 \ddot{f}=6f^2+t \end{equation}
 or
  \begin{equation}
 \ddot{F}=6F^2+t. \end{equation} \\
 c) $F_{IC}(R)$ - model:
   \begin{equation}
 f_{aa}=6f^2+a \end{equation}
 or
  \begin{equation}
 F_{aa}=6F^2+a. \end{equation} \\
 d) $F_{ID}(R)$ - model:
   \begin{equation}
 f_{NN}=6f^2+N \end{equation}
 or
  \begin{equation}
 F_{NN}=6F^2+N. \end{equation} \\
 e) $F_{IE}(R)$ - model:
   \begin{equation}
 f_{HH}=6f^2+H \end{equation}
 or
  \begin{equation}
 F_{HH}=6F^2+H. \end{equation} \\
 2) {\bf $F_{II}(R)$ - model.}\\ a) $F_{IIA}(R)$ - model:
 \begin{equation}
 f^{''}=2f^3+\tilde{R}f+\alpha\end{equation}
 or
  \begin{equation}
 F^{''}=2F^3+\tilde{R}F+\alpha.\end{equation}
 b) $F_{IIB}(R)$ - models:
  \begin{equation}
 \ddot{f}=2f^3+tf+\alpha
  \end{equation}
 or
  \begin{equation}
 \ddot{F}=2F^3+tF+\alpha.
  \end{equation}
   c) $F_{IIC}(R)$ - models:
  \begin{equation}
 f_{aa}=2f^3+af+\alpha
  \end{equation}
 or
  \begin{equation}
 F_{NN}=2F^3+NF+\alpha.
  \end{equation}
 \\ d) $F_{IID}(R)$ - models:
  \begin{equation}
 f_{NN}=2f^3+Nf+\alpha
  \end{equation}
 or
  \begin{equation}
 F_{NN}=2F^3+NF+\alpha.
  \end{equation}
 \\ e) $F_{IIE}(R)$ - models:
  \begin{equation}
 f_{HH}=2f^3+Hf+\alpha
  \end{equation}
 or
  \begin{equation}
 F_{HH}=2F^3+HF+\alpha.
  \end{equation}
  \\
 3) {\bf $F_{III}(R)$ - models.} \\a) $F_{IIIA}(R)$ - model:
 \begin{equation}
 f^{''}=\frac{1}{f}f^{'2}-\frac{1}{\tilde{R}}(f^{'}-\alpha f^2-\beta)+\gamma f^3+\frac{\delta}{f}\end{equation}
 or
  \begin{equation}
 F^{''}=\frac{1}{F}F^{'2}-\frac{1}{\tilde{R}}(F^{'}-\alpha F^2-\beta)+\gamma F^3+\frac{\delta}{F}.\end{equation}
  b) $F_{IIIB}(R)$ - model:
   \begin{equation}
 \ddot{f}=\frac{1}{f}\dot{f}^{2}-\frac{1}{t}(\dot{f}-\alpha f^2-\beta)+\gamma f^3+\frac{\delta}{f}
  \end{equation}
 or
  \begin{equation}
 \ddot{F}=\frac{1}{F}\dot{F}^{2}-\frac{1}{t}(\dot{F}-\alpha F^2-\beta)+\gamma F^3+\frac{\delta}{F}.
  \end{equation}
   c) $F_{IIIC}(R)$ - model:
   \begin{equation}
 f_{aa}=\frac{1}{f}f_{a}^{2}-\frac{1}{a}(f_{a}-\alpha f^2-\beta)+\gamma f^3+\frac{\delta}{f}
  \end{equation}
 or
  \begin{equation}
 F_{aa}=\frac{1}{F}F_{a}^{2}-\frac{1}{a}(F_{a}-\alpha F^2-\beta)+\gamma F^3+\frac{\delta}{F}.
  \end{equation}
 \\
   d) $F_{IIID}(R)$ - model:
   \begin{equation}
 f_{NN}=\frac{1}{f}f_{N}^{2}-\frac{1}{N}(f_{N}-\alpha f^2-\beta)+\gamma f^3+\frac{\delta}{f}
  \end{equation}
 or
  \begin{equation}
 F_{NN}=\frac{1}{F}F_{N}^{2}-\frac{1}{N}(F_{a}-\alpha F^2-\beta)+\gamma F^3+\frac{\delta}{F}.
  \end{equation}
  \\
   e) $F_{IIIE}(R)$ - model:
   \begin{equation}
 f_{HH}=\frac{1}{f}f_{H}^{2}-\frac{1}{H}(f_{H}-\alpha f^2-\beta)+\gamma f^3+\frac{\delta}{f}
  \end{equation}
 or
  \begin{equation}
 F_{HH}=\frac{1}{F}F_{H}^{2}-\frac{1}{H}(F_{a}-\alpha F^2-\beta)+\gamma F^3+\frac{\delta}{F}.
  \end{equation}
 \\
 4) {\bf $F_{IV}(R)$ - models.} \\a) $F_{IVA}(R)$ - model:
 \begin{equation}
 f^{''}=\frac{1}{2f}f^{'2}+1.5f^3+4\tilde{R}f^2+2(\tilde{R}^2-\alpha)f+
 \frac{\delta}{f}\end{equation}
 or
  \begin{equation}
 F^{''}=\frac{1}{2F}F^{'2}+1.5F^3+4\tilde{R}F^2+2(\tilde{R}^2-\alpha)F+
 \frac{\delta}{F}.\end{equation}
  b) $F_{IVB}(R)$ - model:
   \begin{equation}
 \ddot{f}=\frac{1}{2f}\dot{f}^{2}+1.5f^3+4tf^2+2(t^2-\alpha)f+
 \frac{\delta}{f}
  \end{equation}
 or
  \begin{equation}
 \ddot{F}=\frac{1}{2F}\dot{F}^{2}+1.5F^3+4tF^2+2(t^2-\alpha)F+
 \frac{\delta}{F}.
  \end{equation}
   c) $F_{IVC}(R)$ - model:
   \begin{equation}
 f_{aa}=\frac{1}{2f}f_{a}^{2}+1.5f^3+4af^2+2(a^2-\alpha)f+
 \frac{\delta}{f}
  \end{equation}
 or
  \begin{equation}
 F_{aa}=\frac{1}{2F}F_{a}^{2}+1.5F^3+4aF^2+2(a^2-\alpha)F+
 \frac{\delta}{F}.
  \end{equation}
    d) $F_{IVD}(R)$ - model:
   \begin{equation}
 f_{NN}=\frac{1}{2f}f_{N}^{2}+1.5f^3+4Nf^2+2(N^2-\alpha)f+
 \frac{\delta}{f}
  \end{equation}
 or
  \begin{equation}
 F_{NN}=\frac{1}{2F}F_{N}^{2}+1.5F^3+4NF^2+2(N^2-\alpha)F+
 \frac{\delta}{F}.
  \end{equation}
    e) $F_{IVE}(R)$ - model:
   \begin{equation}
 f_{HH}=\frac{1}{2f}f_{H}^{2}+1.5f^3+4Hf^2+2(H^2-\alpha)f+
 \frac{\delta}{f}
  \end{equation}
 or
  \begin{equation}
 F_{HH}=\frac{1}{2F}F_{H}^{2}+1.5F^3+4HF^2+2(H^2-\alpha)F+
 \frac{\delta}{F}.
  \end{equation}
 \\
 5) {\bf $F_{V}(R)$ - models.} \\a) $F_{VA}(R)$ - model:
 \begin{equation}
 f^{''}=(\frac{1}{2f}+\frac{1}{f-1})f^{'2}
 -\frac{1}{\tilde{R}}(f^{'}-\gamma  f)+\tilde{R}^{-2}(f-1)^2(\alpha f+\beta f^{-1})
 +\frac{\delta f(f+1)}{f-1}\end{equation}
 or
  \begin{equation}
 F^{''}=(\frac{1}{2F}+\frac{1}{F-1})F^{'2}
 -\frac{1}{\tilde{R}}(F^{'}-\gamma  F)+\tilde{R}^{-2}(F-1)^2(\alpha F+\beta F^{-1})
 +\frac{\delta F(F+1)}{F-1}.\end{equation}
 b) $F_{VB}(R)$ - model:
   \begin{equation}
 \ddot{f}=(\frac{1}{2f}+\frac{1}{f-1})\dot{f}^{2}
 -\frac{1}{t}(\dot{f}-\gamma  f)+t^{-2}(f-1)^2(\alpha f+\beta f^{-1})
 +\frac{\delta f(f+1)}{f-1}
  \end{equation}
 or
  \begin{equation}
 \ddot{F}=(\frac{1}{2F}+\frac{1}{F-1})\dot{F}^{2}
 -\frac{1}{t}(\dot{F}-\gamma  F)+t^{-2}(F-1)^2(\alpha F+\beta F^{-1})
 +\frac{\delta F(F+1)}{F-1}.
  \end{equation}
   c) $F_{VC}(R)$ - model:
   \begin{equation}
 f_{aa}=(\frac{1}{2f}+\frac{1}{f-1})f_{a}^{2}
 -\frac{1}{a}(f_{a}-\gamma  f)+a^{-2}(f-1)^2(\alpha f+\beta f^{-1})
 +\frac{\delta f(f+1)}{f-1}
  \end{equation}
 or
  \begin{equation}
 F_{aa}=(\frac{1}{2F}+\frac{1}{F-1})F_{a}^{2}
 -\frac{1}{a}(F_{a}-\gamma  F)+a^{-2}(F-1)^2(\alpha F+\beta F^{-1})
 +\frac{\delta F(F+1)}{F-1}.
  \end{equation}
    d) $F_{VD}(R)$ - model:
   \begin{equation}
 f_{NN}=(\frac{1}{2f}+\frac{1}{f-1})f_{N}^{2}
 -\frac{1}{N}(f_{N}-\gamma  f)+a^{-2}(f-1)^2(\alpha f+\beta f^{-1})
 +\frac{\delta f(f+1)}{f-1}
  \end{equation}
 or
  \begin{equation}
 F_{NN}=(\frac{1}{2F}+\frac{1}{F-1})F_{N}^{2}
 -\frac{1}{N}(F_{N}-\gamma  F)+N^{-2}(F-1)^2(\alpha F+\beta F^{-1})
 +\frac{\delta F(F+1)}{F-1}.
  \end{equation}
   e) $F_{VE}(R)$ - model:
   \begin{equation}
 f_{HH}=(\frac{1}{2f}+\frac{1}{f-1})f_{H}^{2}
 -\frac{1}{H}(f_{H}-\gamma  f)+H^{-2}(f-1)^2(\alpha f+\beta f^{-1})
 +\frac{\delta f(f+1)}{f-1}
  \end{equation}
 or
  \begin{equation}
 F_{HH}=(\frac{1}{2F}+\frac{1}{F-1})F_{H}^{2}
 -\frac{1}{H}(F_{H}-\gamma  F)+H^{-2}(F-1)^2(\alpha F+\beta F^{-1})
 +\frac{\delta F(F+1)}{F-1}.
  \end{equation}\\
 6) {\bf $F_{VI}(R)$ - models.} \\a) $F_{VIA}(R)$ - model:
 $$
 f^{''}=0.5\left(\frac{1}{f}+\frac{1}{f-1}+\frac{1}{f-\tilde{R}}\right)f^{'2}
 -\left(\frac{1}{\tilde{R}}+\frac{1}{\tilde{R}-1}+\frac{1}{f-\tilde{R}}\right)f^{'}
 $$\begin{equation}+\tilde{R}^{-2}(\tilde{R}-1)^{-2}f(f-1)(f-\tilde{R})\left[\alpha+ \beta \tilde{R}f^{-2}+\gamma(\tilde{R}-1)(f-1)^{-2}+\delta \tilde{R}(\tilde{R}-1)(f-\tilde{R})^{-2}\right].\end{equation}
 Instead of this equation we can consider the following one
   $$
 F^{''}=0.5\left(\frac{1}{F}+\frac{1}{F-1}+\frac{1}{F-\tilde{R}}\right)F^{'2}
 -\left(\frac{1}{\tilde{R}}+\frac{1}{\tilde{R}-1}+\frac{1}{F-\tilde{R}}\right)F^{'}
 $$\begin{equation}+\tilde{R}^{-2}(\tilde{R}-1)^{-2}F(F-1)(F-\tilde{R})\left[\alpha+ \beta \tilde{R}F^{-2}+\gamma(\tilde{R}-1)(F-1)^{-2}+\delta \tilde{R}(\tilde{R}-1)(F-\tilde{R})^{-2}\right].\end{equation}
 b) $F_{VIB}(R)$ - model: $$
\ddot{f}=0.5\left(\frac{1}{f}+\frac{1}{f-1}+\frac{1}{f-t}\right)\dot{f}^{2}
 -\left(\frac{1}{t}+\frac{1}{t-1}+\frac{1}{f-t}\right)\dot{f}
 $$
  \begin{equation}+t^{-2}(t-1)^{-2}f(f-1)(f-t)\left[\alpha+ \beta tf^{-2}+\gamma(t-1)(f-1)^{-2}+\delta t(t-1)(f-t)^{-2}\right].
  \end{equation}
 and
 $$
 \ddot{F}=0.5\left(\frac{1}{F}+\frac{1}{F-1}+\frac{1}{F-t}\right)\dot{F}^{2}
 -\left(\frac{1}{t}+\frac{1}{t-1}+\frac{1}{F-t}\right)\dot{F}
 $$
  \begin{equation}+t^{-2}(t-1)^{-2}F(F-1)(F-t)\left[\alpha+ \beta tF^{-2}+\gamma(t-1)(F-1)^{-2}+\delta t(t-1)(F-t)^{-2}\right].
  \end{equation}
  c) $F_{VIC}(R)$ - model: $$
f_{aa}=0.5\left(\frac{1}{f}+\frac{1}{f-1}+\frac{1}{f-a}\right)f_{a}^{2}
 -\left(\frac{1}{a}+\frac{1}{a-1}+\frac{1}{f-a}\right)f_{a}
 $$
  \begin{equation}+a^{-2}(a-1)^{-2}f(f-1)(f-a)\left[\alpha+ \beta af^{-2}+\gamma(a-1)(f-1)^{-2}+\delta a(a-1)(f-a)^{-2}\right].
  \end{equation}
 and
 $$
 F_{aa}=0.5\left(\frac{1}{F}+\frac{1}{F-1}+\frac{1}{F-a}\right)F_{a}^{2}
 -\left(\frac{1}{a}+\frac{1}{a-1}+\frac{1}{F-a}\right)F_{a}
 $$
  \begin{equation}+a^{-2}(a-1)^{-2}F(F-1)(F-a)\left[\alpha+ \beta aF^{-2}+\gamma(a-1)(F-1)^{-2}+\delta a(a-1)(F-a)^{-2}\right].
  \end{equation}
   d) $F_{VID}(R)$ - model: $$
f_{NN}=0.5\left(\frac{1}{f}+\frac{1}{f-1}+\frac{1}{f-N}\right)f_{N}^{2}
 -\left(\frac{1}{N}+\frac{1}{N-1}+\frac{1}{f-N}\right)f_{N}
 $$
  \begin{equation}+N^{-2}(N-1)^{-2}f(f-1)(f-N)\left[\alpha+ \beta Nf^{-2}+\gamma(N-1)(f-1)^{-2}+\delta N(N-1)(f-N)^{-2}\right].
  \end{equation}
 and
 $$
 F_{NN}=0.5\left(\frac{1}{F}+\frac{1}{F-1}+\frac{1}{F-N}\right)F_{N}^{2}
 -\left(\frac{1}{N}+\frac{1}{N-1}+\frac{1}{F-N}\right)F_{N}
 $$
  \begin{equation}+N^{-2}(N-1)^{-2}F(F-1)(F-N)\left[\alpha+ \beta NF^{-2}+\gamma(N-1)(F-1)^{-2}+\delta N(N-1)(F-N)^{-2}\right].
  \end{equation}
    e) $F_{VIE}(R)$ - model: $$
f_{HH}=0.5\left(\frac{1}{f}+\frac{1}{f-1}+\frac{1}{f-H}\right)f_{H}^{2}
 -\left(\frac{1}{H}+\frac{1}{H-1}+\frac{1}{f-H}\right)f_{H}
 $$
  \begin{equation}+H^{-2}(H-1)^{-2}f(f-1)(f-H)\left[\alpha+ \beta Hf^{-2}+\gamma(H-1)(f-1)^{-2}+\delta H(H-1)(f-H)^{-2}\right].
  \end{equation}
 and
 $$
 F_{HH}=0.5\left(\frac{1}{F}+\frac{1}{F-1}+\frac{1}{F-H}\right)F_{H}^{2}
 -\left(\frac{1}{H}+\frac{1}{H-1}+\frac{1}{F-H}\right)F_{H}
 $$
  \begin{equation}+H^{-2}(H-1)^{-2}F(F-1)(F-H)\left[\alpha+ \beta HF^{-2}+\gamma(H-1)(F-1)^{-2}+\delta H(H-1)(F-H)^{-2}\right].
  \end{equation}
  So we  presented  new 30   HL cosmological $F_{J}(R) $  models  $(J=I, II, III, IV, V, VI)$ which are integrable due to  integrability  of Painlev$\acute{e}$ equations. 
  
  {\bf Exact solutions of integrable HL models}.   All above constructed integrable HL  models   admit (may be infinity number) exact solutions. Let us here present some of them [for simplicity, we give just some particular solutions for the sister function $f(\tilde{R})=F(\tilde{R})-\tilde{R}$ and only for some models] (see e.g. \cite{Ablowitz}).
  
 1) \textit{The $F_{IIA}(R)$ - model.}  
 
 i) As our first example, we  consider   the $F_{IIA}(R)$ - model (3.17). Let us present its some solutions. For example, this model  has the following particular solutions:
 \begin{eqnarray}
  f(\tilde{R})&\equiv &f(\tilde{R};1.5)=\psi-(2\psi^2+\tilde{R})^{-1},\\
 f(\tilde{R})&\equiv &f(\tilde{R};1)=-\frac{1}{\tilde{R}}, \\ 
  f(\tilde{R})&\equiv &f(\tilde{R};2)=\frac{1}{R}-\frac{3\tilde{R}^2}{\tilde{R}^3+4},\\
  f(\tilde{R})&\equiv& f(\tilde{R};3)=\frac{3\tilde{R}^2}{\tilde{R}^3+4}-\frac{6\tilde{R}^2(\tilde{R}^3+10)}{\tilde{R}^6+20\tilde{R}^3-80},\\
f(\tilde{R})&\equiv& f(\tilde{R};4)=-\frac{1}{\tilde{R}}+\frac{6\tilde{R}^2(\tilde{R}^3+10)}{\tilde{R}^6+20\tilde{R}^3-80}-\frac{9\tilde{R}^5(\tilde{R}^3+40)}{\tilde{R}^9+60\tilde{R}^6+11200},\\
 f(\tilde{R})&\equiv& f(\tilde{R};0.5\epsilon)=-\epsilon\psi
 \end{eqnarray}
and so on. Here 
 \begin{equation}
\psi=(\ln{\phi})_{\tilde{R}}, \quad  \phi(\tilde{R})=C_1Ai(-2^{-1/3}\tilde{R})+C_2Bi{(-2^{-1/3}\tilde{R})},
 \end{equation}
 $C_i=consts$ and $Ai(x), Bi(x)$ are Airy functions. 
 
 ii) Similarly, for the  $F_{IIB}(R)$ - model (3.19) we have the following particular solutions
  \begin{eqnarray}
  f(\tilde{R})&\equiv &f(t;1.5)=\psi-(2\psi^2+t)^{-1},\\
 f(\tilde{R})&\equiv &f(t;1)=-\frac{1}{t}, \\ 
  f(\tilde{R})&\equiv &f(t;2)=\frac{1}{R}-\frac{3t^2}{t^3+4},\\
  f(\tilde{R})&\equiv& f(t;3)=\frac{3t^2}{t^3+4}-\frac{6t^2(t^3+10)}{t^6+20t^3-80},\\
f(\tilde{R})&\equiv& f(t;4)=-\frac{1}{t}+\frac{6t^2(t^3+10)}{t^6+20t^3-80}-\frac{9t^5(t^3+40)}{t^9+60t^6+11200},\\
 f(\tilde{R})&\equiv& f(t;0.5\epsilon)=-\epsilon\psi
 \end{eqnarray}
and so on. Here 
 \begin{equation}
\psi=(\ln{\phi})_{t}, \quad  \phi(t)=C_1Ai(-2^{-1/3}t)+C_2Bi{(-2^{-1/3}t)}.\end{equation}

 2) \textit{The $F_{IIIA}(R)$ - model.} 
 
 i)  Our next example is the $F_{IIIA}(R)$ - model (3.27). It has the following particular solutions:
 \begin{eqnarray}
  f(\tilde{R})&\equiv &f(\tilde{R};\nu_1,0,0,-\nu_1\nu_2^3)=\nu_2\sqrt[3]{\tilde{R}},\\
 f(\tilde{R})&\equiv &f(\tilde{R};0,-2\nu_1,0,4\nu_1\nu_2-\nu_3^2)=\tilde{R}[\ln{(\tilde{R}^{\sqrt{\nu_1}})^2}+\ln{(\tilde{R}^{\nu_3}e^{\nu_2})}], \\ 
  f(\tilde{R})&\equiv &f(\tilde{R};-\nu_1^2\nu_2,0,\nu_1^2(\nu_2^2-\nu_3\nu_4),0)=\frac{\tilde{R}^{\nu_1-1}}{\nu_3\tilde{R}^{2\nu_1}+\nu_2\tilde{R}^{\nu_1}+\nu_4},\\
  f(\tilde{R})&\equiv& f(\tilde{R};2\nu_1+3,-2\nu_1+1,1,-1)=\frac{\tilde{R}+\nu_1}{\tilde{R}+\nu_1+1},\\
f(\tilde{R})&\equiv& f(\tilde{R};\nu_1,-\nu_1\nu_2^2,\nu_3,-\nu_3\nu_2^4)=\nu_2,\\
 f(\tilde{R})&\equiv& f=-\epsilon_1(\ln{\varphi})_{\tilde{R}}
 \end{eqnarray}
and so on. Here 
 \begin{equation}
\varphi(\tilde{R})=\tilde{R}^{\nu}[C_1J_\nu(\sqrt{\epsilon_1\epsilon_2}\tilde{R})+C_2Y_\nu(\sqrt{\epsilon_1\epsilon_2}\tilde{R})], \quad (\epsilon_i, C_i=consts),\end{equation}
 and $J_\nu(x), Y_\nu(x)$ are Bessel  functions.
 
 ii) The corresponding particular solutions of the  $F_{IIIB}(R)$ - model (3.29) are given by
 \begin{eqnarray}
  f(\tilde{R})&\equiv &f(t;\nu_1,0,0,-\nu_1\nu_2^3)=\nu_2\sqrt[3]{t},\\
 f(\tilde{R})&\equiv &f(t;0,-2\nu_1,0,4\nu_1\nu_2-\nu_3^2)=t[\ln{(t^{\sqrt{\nu_1}})^2}+\ln{(t^{\nu_3}e^{\nu_2})}], \\ 
  f(\tilde{R})&\equiv &f(t;-\nu_1^2\nu_2,0,\nu_1^2(\nu_2^2-\nu_3\nu_4),0)=\frac{t^{\nu_1-1}}{\nu_3t^{2\nu_1}+\nu_2t^{\nu_1}+\nu_4},\\
  f(\tilde{R})&\equiv& f(t;2\nu_1+3,-2\nu_1+1,1,-1)=\frac{t+\nu_1}{t+\nu_1+1},\\
f(t)&\equiv& f(t;\nu_1,-\nu_1\nu_2^2,\nu_3,-\nu_3\nu_2^4)=\nu_2,\\
 f(\tilde{R})&\equiv& f=-\epsilon_1(\ln{\varphi})_{t}
 \end{eqnarray}
and so on. Here 
 \begin{equation}
\varphi(t)=t^{\nu}[C_1J_\nu(\sqrt{\epsilon_1\epsilon_2}t)+C_2Y_\nu(\sqrt{\epsilon_1\epsilon_2}t)].\end{equation}
  
  Similarly we can present the exact solutions of the other models.
 \subsection{Nonintegrable models}
Let us here present   some known and new  HL models induced by  some  ODEs. These ODEs are   nonintegrable so that the corresponding HL 
cosmological models are nonintegrable. Consider examples. 
 
 i) We start from models induced by   the hypergeometric differential equation. We write here 5 versions of this model.\\
R-version:
 \begin{equation}
\tilde{R}(1-\tilde{R})f^{''}+[c-(\alpha+b+1)\tilde{R}]f^{'}-\alpha bf=0. \end{equation}
It has the solution $f(\tilde{R})=\,_2F(\alpha,b; c; \tilde{R})$ which is the hypergeometric function. \\
t-version: 
 \begin{equation}
t(1-t)\ddot{f}+[c-(\alpha+b+1)t]\dot{f}-\alpha bf=0 \end{equation}
with the solution $f(t)=\,_2F(\alpha,b; c; t)$.
 \\
a-version: 
 \begin{equation}
a(1-a)f_{aa}+[c-(\alpha+b+1)t]f_{a}-\alpha bf=0 \end{equation}
with the solution $f(a)=\,_2F(\alpha,b; c; a)$.\\
N-version: 
 \begin{equation}
N(1-N)f_{NN}+[c-(\alpha+b+1)t]f_{N}-\alpha bf=0 \end{equation}
with the solution $f(N)=\,_2F(\alpha,b; c; N)$.
\\
H-version: 
 \begin{equation}
H(1-H)f_{HH}+[c-(\alpha+b+1)t]f_{H}-\alpha bf=0 \end{equation}
with the solution $f(H)=\,_2F(\alpha,b; c; H)$.

ii)  Another example is the case when $f(\tilde{R})$ satisfies the Pinney equation
 \begin{equation}
f^{''}+\xi_{1}(\tilde{R})f+\frac{\xi_{2}(\tilde{R})}{f^3}=0. \end{equation}
 If $\xi_1=1,\quad \xi_2=\kappa=const$, this equation has the following solution (see e.g. \cite{MR})
 \begin{equation}
f(\tilde{R})=\cos^2\tilde{R}+\kappa^2\sin^2\tilde{R}. \end{equation}
 Its  "$t$-form" is $f(t)=\cos^2t+\kappa^2\sin^2t$ which is  the solution of the  Pinney equation  \begin{equation}
\ddot{f}+\xi_{1}(t)f+\frac{\xi_{2}(t)}{f^3}=0 \end{equation}
 as $\xi_1=1,\quad \xi_2=\kappa=const$.
 
 iii) Let us we present one more example. Let the function $f$ satisfies the equation
  \begin{equation}
f^{''}=6f^2-0.5g_2, \quad (g_2=const)\end{equation}
  or
    \begin{equation}
\ddot{f}=6f^2-0.5g_2.\end{equation}These equations admit the following solutions
   \begin{equation}
f(\tilde{R})=\wp(\tilde{R}) \end{equation}
and
   \begin{equation}
f(\tilde{R})\equiv f(t)=\wp(t), \end{equation}
where  $\wp(\tilde{R})$ and $\wp(t)$ are  the Weierstrass elliptic functions. 
 
 	Finally we would like to note that the above presented HL models, in particular, have some solutions which describe the accelerated expansion of the universe. 

 	\section{Cosmological solutions}  It is important find exact solutions of HL gravity models (see e.g. \cite{Elizalde}-\cite{Saridakis}).  In our case, all above presented HL models admit exact solutions. It is important that some of these solutions describe accelerated expansion of the universe.  Let us present  some cosmological solutions of some  above presented HL models. As an example, consider the P$_{II}$-models that is  the equations (2.27)-(2.31). These equations  have the following particular solutions (see e.g. \cite{Ablowitz}-\cite{MR}). 
 	
i) P$_{IIA}$ - model (2.27):
\begin{eqnarray}
  N(t)&\equiv &N(t;\nu_1,0,0,-\nu_1\nu_2^3)=\nu_2\sqrt[3]{t},\\
 N(t)&\equiv &N(t;0,-2\nu_1,0,4\nu_1\nu_2-\nu_3^2)=t[(\ln{t^{\sqrt{\nu_1}})^2}+\ln{(t^{\nu_3}e^{\nu_2})}], \\ 
  N(t)&\equiv &N(t;-\nu_1^2\nu_2,0,\nu_1^2(\nu_2^2-\nu_3\nu_4),0)=\frac{t^{\nu_1-1}}{\nu_3t^{2\nu_1}+\nu_2t^{\nu_1}+\nu_4},\\
  N(t)&\equiv& N(t;2\nu_1+3,-2\nu_1+1,1,-1)=\frac{t+\nu_1}{t+\nu_1+1},\\
N(t)&\equiv& N(t;\nu_1,-\nu_1\nu_2^2,\nu_3,-\nu_3\nu_2^4)=\nu_2,\\
 N(t)&\equiv& N=-\epsilon_1(\ln{\varphi})_{t}.
 \end{eqnarray}
 Hence we get the corresponding expressions for the scale factor $a(t)$. We have
 \begin{eqnarray}
  a(t)&\equiv &a(t;\nu_1,0,0,-\nu_1\nu_2^3)=e^{\nu_2\sqrt[3]{t}},\\
 a(t)&\equiv &a(t;0,-2\nu_1,0,4\nu_1\nu_2-\nu_3^2)=t^{\nu_3t}e^{\nu_2t}e^{(\ln{t^{\sqrt{\nu_1t}})^2}}, \\ 
  a(t)&\equiv &a(t;-\nu_1^2\nu_2,0,\nu_1^2(\nu_2^2-\nu_3\nu_4),0)=e^{\frac{t^{\nu_1-1}}{\nu_3t^{2\nu_1}+\nu_2t^{\nu_1}+\nu_4}},\\
  a(t)&\equiv& a(t;2\nu_1+3,-2\nu_1+1,1,-1)=e^{\frac{t+\nu_1}{t+\nu_1+1}},\\
a(t)&\equiv& a(t;\nu_1,-\nu_1\nu_2^2,\nu_3,-\nu_3\nu_2^4)=e^{\nu_2},\\
 a(t)&\equiv& a=e^{-\epsilon_1(\ln{\varphi})_{t}}.
 \end{eqnarray}
ii)  P$_{IIB}$ - model (2.28):
\begin{eqnarray}
  a(t)&\equiv &a(t;\nu_1,0,0,-\nu_1\nu_2^3)=\nu_2\sqrt[3]{t},\\
 a(t)&\equiv &a(t;0,-2\nu_1,0,4\nu_1\nu_2-\nu_3^2)=t[\ln{(t^{\sqrt{\nu_1}})^2}+\ln{(t^{\nu_3}e^{\nu_2})}], \\ 
  a(t)&\equiv &a(t;-\nu_1^2\nu_2,0,\nu_1^2(\nu_2^2-\nu_3\nu_4),0)=\frac{t^{\nu_1-1}}{\nu_3t^{2\nu_1}+\nu_2t^{\nu_1}+\nu_4},\\
  a(t)&\equiv& a(t;2\nu_1+3,-2\nu_1+1,1,-1)=\frac{t+\nu_1}{t+\nu_1+1},\\
a(t)&\equiv& a(t;\nu_1,-\nu_1\nu_2^2,\nu_3,-\nu_3\nu_2^4)=\nu_2,\\
 a(t)&\equiv& a=-\epsilon_1(\ln{\varphi})_{t}.
 \end{eqnarray}
  
 iii)  P$_{IIC}$ - model (2.29):
 \begin{eqnarray}
  H(t)&\equiv &H(t;\nu_1,0,0,-\nu_1\nu_2^3)=\nu_2\sqrt[3]{t},\\
 H(t)&\equiv &H(t;0,-2\nu_1,0,4\nu_1\nu_2-\nu_3^2)=t[\ln{(t^{\sqrt{\nu_1}})^2}+\ln{(t^{\nu_3}e^{\nu_2})}], \\ 
  H(t)&\equiv &H(t;-\nu_1^2\nu_2,0,\nu_1^2(\nu_2^2-\nu_3\nu_4),0)=\frac{t^{\nu_1-1}}{\nu_3t^{2\nu_1}+\nu_2t^{\nu_1}+\nu_4},\\
  H(t)&\equiv& H(t;2\nu_1+3,-2\nu_1+1,1,-1)=\frac{t+\nu_1}{t+\nu_1+1},\\
H(t)&\equiv& H(t;\nu_1,-\nu_1\nu_2^2,\nu_3,-\nu_3\nu_2^4)=\nu_2,\\
 H(t)&\equiv& H=-\epsilon_1(\ln{\varphi})_{t}
 \end{eqnarray}
 and so on. These exact solutions correspond to the different cosmologies. Some of them describe the accelerated and decelarated phases  of the Universe. Finally we note that similarly we can present exact solutions of the other HL models constructed in the previous sections. 
\section{Conclusion}

In this work,  a new class integrable and nonintegrable cosmological models of the  Ho$\check{r}$ava-Lifshitz  gravity  were proposed. For some of them, exact solutions are presented. To construct integrable models we use the well-known integrable systems, namely, Painleve equations. Then these results extend for the F(R) Ho$\check{r}$ava-Lifshitz  gravity theory case. In particular, several integrable cosmological models of this modified gravity theory were constructed in the explicit form.


\begin{thebibliography}{99}
   \bibitem{Ho} Horava P. 
   Phys. Rev. D, {\bf 79}, 084008  (2009) [arXiv:0901.3775]
   \bibitem{Chaichian} Chaichian M.,   Nojiri S.,  Odintsov S.D.,  Oksanen M., Tureanu A. 
   Class. Quantum Grav., {\bf 27}, 185021  (2010) [arXiv:1001.4102]
     \bibitem{Odin1} Nojiri S., Odintsov S.D. \textit{Unified cosmic history in modified gravity: from F(R) theory to Lorentz non-invariant models}, Physics Reports, (2011) [arXiv:1011.0544].
  \bibitem{Elizalde} Elizalde E.,   Nojiri S.,  Odintsov S.D., Saez-Gomez. 
   Eur. Phys. J. C, {\bf 70}, 351  (2010) [arXiv:1006.3387]
   \bibitem{Nojiri11}   Nojiri S.,  Odintsov S.D., Saez-Gomez D. 
    Phys. Lett. B, {\bf 681}, 74  (2009) [arXiv:0908.1269]
    
     \bibitem{Saez1}    Saez-Gomez D. Phys. Rev.D, {\bf 83}, 064040 (2011) [arXiv:1011.2090]
   \bibitem{Kluson}  Kluson J.,  Nojiri S.,  Odintsov S.D., Saez-Gomez D. \textit{U(1) Invariant F(R) Horava-Lifshitz Gravity}, 
    [arXiv:1012.0473]
    \bibitem{Saez2}  Saez-Gomez D.  \textit{Cosmological solutions in F(R) Horava-Lifshitz gravity}, [arXiv:1012.4605]
    \bibitem{Saridakis} Saridakis E.N. \textit{Aspects of Horava-Lifshitz cosmology}, [arXiv:1101.0300];\\
    Ali A.,  Dutta S.,  Saridakis E.N.,  Sen A.A. \textit{Horava-Lifshitz cosmology with generalized Chaplygin gas}, [arXiv:1004.2474];\\
      Jamil M.,  Saridakis E.N., Setare M. R. JCAP, {\bf  1011}, 032 (2010),[arXiv:1003.0876]


         \bibitem{Ablowitz} Ablowitz M. J, Clarkson P. A. \textit{Solitons, nonlinear evolution equations and inverse scattering}, London Mathematical Society Lecture Note Series, {\bf 149}, Cambridge University Press, 1991;\\
     Clarkson P.A. \textit{Painlev$\acute{e}$  transcendents}, in NIST Handbook of Mathematical Functions, Cambridge University Press, 2010.
    \bibitem{MR}  Esmakhanova K., Myrzakulov N., Nugmanova G.,  Chechin L.,   Myrzakulov R. \textit{Integrable and nonintegrable FRW cosmological models induced by some second-order ordinary differential equations}, [arXiv:1104.3705]
    \end{thebibliography}
 \end{document}